\documentclass[a4paper,fleqn]{cas-sc}
\usepackage[numbers,sort&compress]{natbib}
\usepackage{amsmath,amssymb,bm,mathtools}
\usepackage{graphicx}
\usepackage{booktabs}
\usepackage{longtable}
\usepackage{multirow}
\usepackage{array}
\usepackage{float}
\usepackage{makecell}
\usepackage{caption}
\usepackage[section]{placeins}
\usepackage{hyperref}
\graphicspath{{./}}
\newcommand{\Eq}[1]{Eq.(\ref{#1})}
\newcommand{\Fig}[1]{Fig.(\ref{#1})}
\newcommand{\Tab}[1]{Table \ref{#1}}
\newcommand{\SMSec}[1]{SM Section~#1}

\newcommand{\bU}{\bm{U}}

\newcommand{\dd}{\mathrm{d}}

\begin{document}
\let\WriteBookmarks\relax

\shorttitle{LSTM-PINN for Steady Electrothermal Multiphysics}
\shortauthors{Y.Zhou et~al.}

\title[mode=title]{LSTM-PINN for Steady-State Electrothermal Transport: Preserving Multi-Field Consistency in Strongly Coupled Heat and Fluid Flow}

\author[1]{Yuqing Zhou}[orcid=0009-0005-5038-6604]
\credit{Calculation, data analyzing and manuscript writing}
\fnref{co-first}
\author[1]{Ze Tao}[orcid=0009-0004-0202-3641]
\credit{Calculation, data analyzing, manuscript writing, review and editing}
\fnref{co-first}
\affiliation[1]{organization={Nanophotonics and Biophotonics Key Laboratory of Jilin Province, School of Physics, Changchun University of Science and Technology},
                city={Changchun},
                postcode={130022},
                country={P.R. China}}
\author[2,3]{Hanxuan Wang}[orcid=0000-0003-1830-5913]
\credit{Data analyzing}
\affiliation[2]{organization={Faculty of Chinese Medicine, Macau University of Science and Technology, Taipa},
                city={Macau},
                postcode={999078},
                country={P.R. China}}
\affiliation[3]{organization={Laboratory of Drug Discovery from Natural Resources and Industrialization, Macau University of Science and Technology, Taipa},
                city={Macau},
                postcode={999078},
                country={P.R. China}}
\author[1]{Fujun Liu}[orcid=0000-0002-8573-450X]
\corref{cor}
\credit{Review and Editing}
\ead{fjliu@cust.edu.cn}
\fntext[co-first]{The authors contribute equally to this work.}
\cortext[cor]{Corresponding author}

\begin{abstract}
Steady-state electrothermal systems involve strongly coupled heat transfer, fluid flow, and electric-potential transport, creating severe numerical challenges for standard physics-informed neural networks (PINNs) due to stark disparities in gradient scales and residual stiffnesses across the physical fields. To resolve these multiphysics bottlenecks, we introduce a Long Short-Term Memory PINN (LSTM-PINN) framework that utilizes a depth-recursive memory mechanism to preserve long-range spatial feature dependencies and maintain strict cross-field consistency. The proposed architecture is rigorously evaluated against conventional and attention-based networks across a unified five-field formulation encompassing four complex convective and drag regimes: Boussinesq electrothermal flow, drift-potential gauge-constrained transport, strong buoyancy-coupled convection, and Brinkman--Forchheimer drift. Quantitative and visual analyses demonstrate that LSTM-PINN successfully suppresses non-physical artifacts and structural distortions, yielding the highest thermodynamic fidelity and consistently outperforming state-of-the-art baselines in global error metrics. Ultimately, this memory-enhanced approach provides a highly robust and accurate computational baseline for capturing localized boundary layers and complex energy-momentum feedback in advanced electrothermal energy systems.
\end{abstract}

\begin{highlights}
\item LSTM-PINN resolves strongly coupled steady-state electrothermal transport.
\item Unified five-field PDE captures momentum, heat, and electric transfer.
\item Depth-recursive memory strictly preserves cross-field thermodynamic consistency.
\item Architecture effectively resolves steep boundary layers and non-linear drag.
\end{highlights}

\begin{keywords}
Physics-informed neural network \sep LSTM-PINN \sep Electrothermal coupling \sep Steady-state multiphysics
\end{keywords}

\maketitle
\section{Introduction}
Partial differential equations (PDE) govern momentum transport, pressure redistribution, electric-potential evolution, and thermal diffusion in a wide range of electrothermal systems. When these mechanisms coexist in a single domain, the unknown variables interact intricately through advection, diffusion, body-force coupling, and constitutive feedback, creating strongly coupled multiphysics problems relevant to thermal storage, liquid cooling, and energy devices \citep{heatreview2025,borehole2025,thermoelectric2025,heatstorage2025,pisanet2026}. Recent studies on battery state estimation and PEM fuel-cell optimization further underscore the engineering value of physics-informed thermal modeling \citep{batteryhybrid2025,batterypiml2026,batterycoest2025,batterysoc2025,pemfc2026}. Physics-informed neural networks provide a natural route for addressing these transport phenomena by embedding governing equations and boundary constraints directly into the objective function. Recent papers in \textit{International Journal of Heat and Mass Transfer} have already extended PINN-based thermal modeling to thin-film evaporation, non-Fourier heat conduction, bubble growth, packed-bed diffusion inversion, and two-phase film boiling \citep{jahanbakhsh2024thinfilm,zheng2024tph,jalili2024bubble,wang2025packedbed,jalili2025filmboiling}. Additional \textit{International Journal of Heat and Mass Transfer} studies further address inverse thermal-wave reconstruction, surface-heat-flux estimation, and anisotropic heat conduction \citep{tang2025thermalwave,cai2025surfaceflux,gao2025rbf}. However, in these complex electrothermal systems, the physical driving forces often operate on vastly different characteristic scales. For instance, the natural convection buoyancy induced by sharp temperature gradients and the Coulomb forces generated by the external electric field typically exhibit extreme disparities in magnitude. Furthermore, the thermal transport pathways and the electric charge distributions are frequently misaligned in space. This physical spatial asynchrony and multi-scale force disparity manifest mathematically as severe residual stiffness and highly imbalanced gradient scales, causing standard feedforward multilayer perceptron backbones to suffer from localized information breakage and non-physical artifacts.

To mitigate these optimization bottlenecks and improve the thermodynamic fidelity of the resolved fields, recent work has pushed network design toward residual attention, loss-attentional weighting, balanced residual decay, dual adaptivity, and gradient-aware stabilization \citep{rba2024,lapinn2024,balanced2025,adual2025,ferrer2024gapinn}. Mixed-depth or nested-basis formulations further enrich the design space for stiff PDE solvers \citep{mdpinn2025}. Concurrently, the broader machine learning literature has expanded its architectural toolkit to better capture complex physical behaviors, employing methods such as domain decomposition, structure-preserving spaces, and neural operators to maintain conservation laws while keeping the highly non-linear optimization tractable \citep{ddd2024,afeinn2025,cfeinn2025,multiscale2024,naspinn2024}. Learnable activations, trainable sinusoidal bases, transformer operators, porous-media neural operators, and high-dimensional PINN analysis further expand the available architectures \citep{laf2025,tsa2025,pinto2025,pinoinfiltration2025,curse2024}.

Within this evolving landscape, the Long Short-Term Memory physics-informed neural network (LSTM-PINN) occupies a highly distinctive position. Instead of relying exclusively on pointwise feedforward feature propagation, this architecture introduces gated memory transport into the network backbone and utilizes recurrent state updates to regulate how intermediate spatial features survive across the network depth. Crucially, this design does not add a physical time variable to the steady-state governing equations. Rather, it incorporates a learnable depth-wise memory mechanism that acts as an active filter for cross-field thermodynamic consistency. While conventional perceptrons iteratively lose macroscopic energy conservation constraints when aggressively fitting steep local gradients, the LSTM's gated memory actively retains and propagates the cross-scale energy-momentum coupling relationships across the network depth. This ensures the fluid momentum simultaneously and coherently responds to both localized thermal buoyancy and electrohydrodynamic drift. This perspective aligns with recent memory-aware structures in steady electrohydrodynamics, building thermal control, interface-dominated multiphysics, and nonlinear heat conduction \citep{tao2025lstm,buildingct2026,xing2025modeling,tao2025analytical}. Related recurrent explorations have also appeared in broader forecasting settings and physics-only training frameworks \citep{tao2025lstm_population,tao2025lnn}.

To rigorously evaluate the capability of this architecture in capturing complex thermal and fluid interactions, we systematically investigate the LSTM-PINN framework across four distinct steady-state electrothermal regimes. We maintain a unified five-field formulation and directly compare the proposed method against state-of-the-art baselines, including a residual-attention model (ResAttn-PINN) \citep{zhou2026residual}, a parallel LSTM network (pLSTM-PINN), and a purely feedforward perceptron (Pure-MLP). This comprehensive evaluation spans a constant-coefficient Boussinesq electrothermal flow, a drift-potential gauge-constrained configuration, a strong buoyancy-coupled transport scenario, and a highly non-linear Brinkman--Forchheimer drift case. By applying the memory-enhanced depth-recursive architecture to these demanding physical problems, we aim to demonstrate its capacity to resolve localized convection phenomena and maintain strict cross-field physical consistency under varying boundary constraints and coupling strengths.

The quantitative and visual evidence from these systematic evaluations reveals a consistent advantage for the memory-enhanced formulation in resolving strongly coupled transport. Across the four analyzed physical configurations, the LSTM-PINN framework achieves the lowest average absolute errors when integrating the five coupled fields, outperforming the competing models in resolving complex global structures and steep local gradients. While all models exhibit sensitivity to scaling in the interface-dominated Brinkman--Forchheimer regime, the absolute metric comparisons confirm the superior numerical stability of the depth-recursive approach in managing intense, localized non-linear dissipation. Consequently, this study demonstrates that the LSTM-PINN backbone delivers an optimal balance between thermodynamic accuracy, optimization robustness, and structural interpretability for steady electrothermal multiphysics, with extensive technical documentation supporting these findings detailed in the accompanying supplementary material (SM).

\section{LSTM-PINN Architecture}

\begin{figure}[t]
    \centering
    \includegraphics[width=0.98\textwidth]{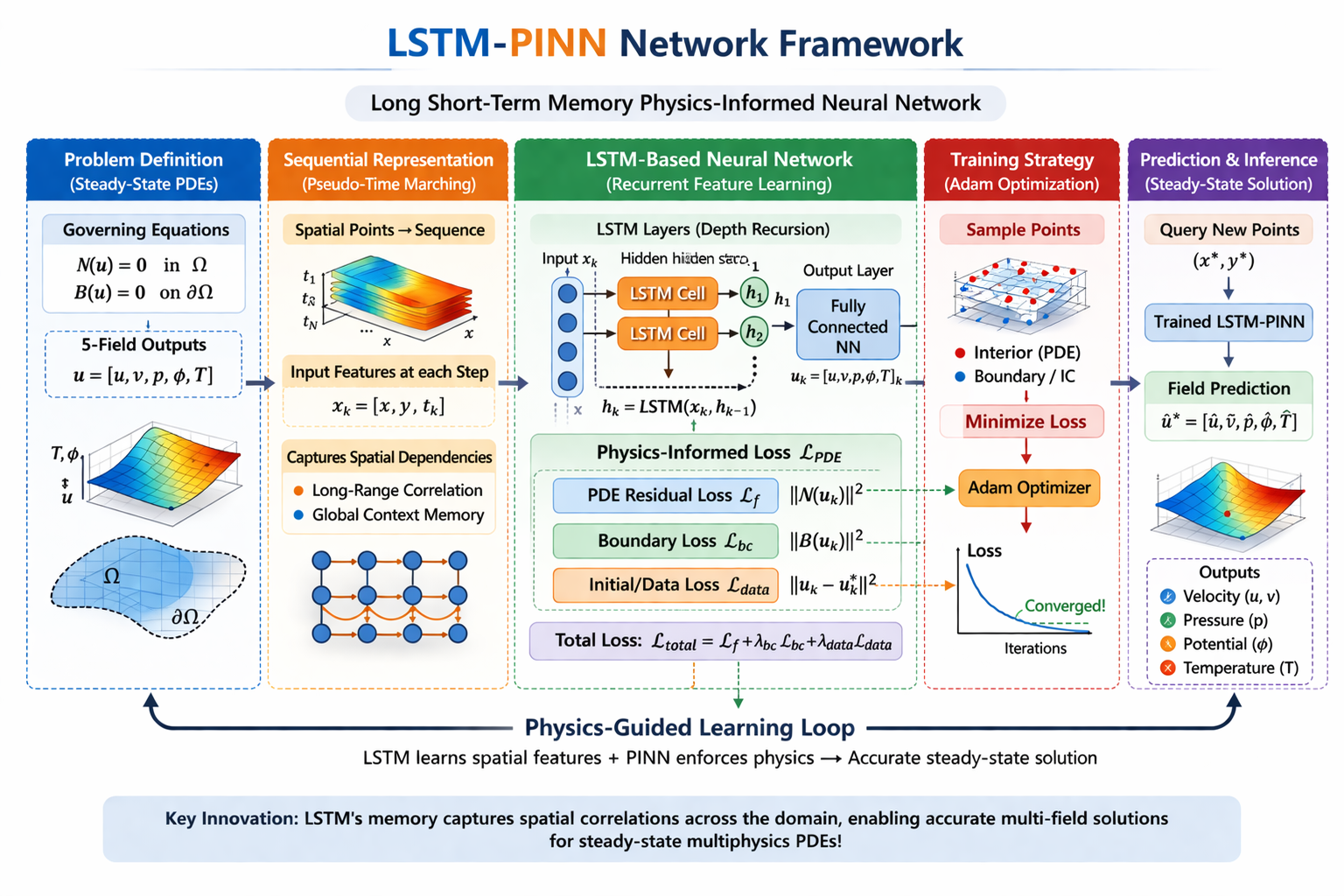}
    \caption{The proposed LSTM-PINN framework for steady-state electrothermal systems. The diagram details the depth-recursive architecture, spatial coordinate encoding, composite physical residual evaluation, and gated memory transport, integrating them into a cohesive workflow tailored to resolve the strongly coupled five-field interactions.}
    \label{fig:lstm_arch}
\end{figure}
\FloatBarrier

The framework illustrated in \Fig{fig:lstm_arch} establishes the computational design of the memory-enhanced solver, translating pointwise spatial coordinates into a thermodynamically consistent multi-field prediction. To process the spatial domain, the network maps the physical coordinates to the solution fields through a depth-recursive sequence of Long Short-Term Memory cells. Mathematically, the continuous mapping from spatial coordinates to the governing physical fields is defined as
\begin{equation}
\widehat{\bU}(x,y;\theta)=\mathcal{F}_{\theta}(x,y)=\left[\hat u(x,y),\hat v(x,y),\hat p(x,y),\hat T(x,y),\hat \phi(x,y)\right]^\top.
\label{eq:lstm_map}
\end{equation}
Prior to entering the recursive structure, the spatial inputs undergo a preprocessing transformation. This initial feature extraction reads
\begin{equation}
\bm{r}^{(0)}=\tanh\!\left(W_{\mathrm{in}}\,\bm{z}+\bm{b}_{\mathrm{in}}\right),
\qquad
\bm{z}=2\,[x,y]^\top-\bm{1},
\label{eq:lstm_inputproj}
\end{equation}
which is initialized with spatial recurrent states $\bm{h}^{(0)}=\bm{r}^{(0)}$ and a zeroed structural cell state $\bm{c}^{(0)}=\bm{0}$. 

Building upon this initialized state, the features are propagated through the network depth rather than advancing through time. For the $\ell$th recursive layer, the spatial feature state update is governed by
\begin{equation}
\left(\bm{h}^{(\ell)},\bm{c}^{(\ell)}\right)=\mathrm{LSTMCell}_{\ell}\!\left(\bm{r}^{(\ell-1)},\left(\bm{h}^{(\ell-1)},\bm{c}^{(\ell-1)}\right)\right),
\qquad
\bm{r}^{(\ell)}=\bm{h}^{(\ell)},
\label{eq:lstm_depthrec}
\end{equation}
for $\ell=1,2,\dots,L$. Within each individual depth cell, the memory update mechanism operates strictly according to
\begin{equation}
\bm{c}_{t}=\bm{f}_{t}\odot\bm{c}_{t-1}+\bm{i}_{t}\odot\widetilde{\bm{c}}_{t},
\qquad
\bm{h}_{t}=\bm{o}_{t}\odot\tanh\!\left(\bm{c}_{t}\right).
\label{eq:lstm_cellupdate}
\end{equation}
In this physical formulation, the forget, input, and output gates act as active filters for cross-field consistency. They continuously regulate how long-range spatial correlations and energy-momentum coupling behaviors are retained, updated, or suppressed as the feature information travels deeper into the network architecture. Once the deep structural representation $\bm{h}^{(L)}$ is established, a subsequent fully connected neural network block projects these refined intermediate features into the final macroscopic variables. 

After the network generates the comprehensive field prediction $\widehat{\bU}$, automatic differentiation evaluates the spatial gradients required by the governing partial differential equations. This step allows the framework to precisely construct the continuity, momentum, energy, and electric-potential residual formulations, which are subsequently minimized alongside the prescribed boundary constraints. Comprehensive details regarding the explicit stacked-gate recursion formulation, the physical interpretation of the memory transport mechanism, and the precise architectural configurations for all evaluated networks are documented in \SMSec{II} of the SM.

\section{Unified PDE Formulation}
We write the steady electrothermal coupled system in one operator form over the unit square domain $\Omega=[0,1]\times[0,1]$. The unknown vector reads
\begin{equation}
\bU(x,y)=\left[u(x,y),v(x,y),p(x,y),\phi(x,y),T(x,y)\right]^\top.
\label{eq:U}
\end{equation}
We denote the two-dimensional gradient by $\nabla=(\partial_x,\partial_y)^\top$ and the Laplacian by $\Delta=\partial_{xx}+\partial_{yy}$.

The steady incompressibility constraint reads
\begin{equation}
R_c(\bU)=\nabla\cdot\bm{u}=\partial_x u+\partial_y v=0,
\label{eq:cont}
\end{equation}
where $\bm{u}=(u,v)^\top$. We write the momentum equations as
\begin{equation}
\bm{R}_{\mathrm{mom}}(\bU)=\rho\,(\bm{u}\cdot\nabla)\bm{u}+\nabla p-\nabla\cdot\!\left(\nu\nabla\bm{u}\right)-\bm{f}_{\mathrm{e}}(\phi,T)=\bm{0},
\label{eq:mom}
\end{equation}
where $\rho$ denotes density, $\nu$ denotes the viscosity-related transport coefficient, and $\bm{f}_{\mathrm{e}}$ denotes the electrothermal body-force contribution.

The electric-potential equation reads
\begin{equation}
R_{\phi}(\bU)=-\nabla\cdot\!\left(\sigma\nabla\phi\right)-s_{\phi}(x,y)=0,
\label{eq:phi_eq}
\end{equation}
where $\sigma$ denotes electric conductivity and $s_{\phi}$ denotes the source term selected by the benchmark. The temperature equation reads
\begin{equation}
R_T(\bU)=-\nabla\cdot\!\left(\alpha\nabla T\right)+\bm{u}\cdot\nabla T-s_T(x,y)=0,
\label{eq:T_eq}
\end{equation}
where $\alpha$ denotes thermal diffusivity and $s_T$ denotes the thermal source term.

We collect the full interior residual in one vector operator,
\begin{equation}
\mathcal{N}(\bU)=\left[R_c,\,R_u,\,R_v,\,R_{\phi},\,R_T\right]^\top=\bm{0},
\label{eq:operator}
\end{equation}
where $R_u$ and $R_v$ denote the two scalar components of \Eq{eq:mom}. This notation lets us keep one PDE template across all four uploaded cases.

The case physics differ through the source terms, the constitutive law, and the auxiliary constraints. Case~1 uses a Boussinesq electrothermal incompressible Navier--Stokes system with no-slip velocity boundaries, mixed thermal and electric boundary conditions, and a zero-mean pressure gauge. Case~2 adds drift-potential coupling and again reconstructs pressure through a global gauge constraint rather than a pointwise anchor. Case~3 retains the five-field electrothermal form but activates buoyancy-sensitive forcing in the vertical momentum equation. Case~4 introduces Brinkman--Forchheimer drag through the nonlinear term $\beta_f |\bm{u}|\bm{u}$ and combines it with drift-potential transport.

For a network prediction $\widehat{\bU}(x,y;\theta)$, we define the interior residual vector by
\begin{equation}
\mathcal{R}_{\Omega}(x,y;\theta)=\mathcal{N}(\widehat{\bU}(x,y;\theta)),
\qquad (x,y)\in \Omega,
\label{eq:res_omega}
\end{equation}
and we define the boundary residual by
\begin{equation}
\mathcal{R}_{\partial\Omega}(x,y;\theta)=\mathcal{B}(\widehat{\bU}(x,y;\theta))-\bm{G}(x,y),
\qquad (x,y)\in \partial\Omega,
\label{eq:res_boundary}
\end{equation}
where $\mathcal{B}$ collects the prescribed boundary operators and $\bm{G}$ collects the corresponding boundary data. In the pressure-gauge cases, we also use the global residual
\begin{equation}
\mathcal{R}_{\mathrm{g}}(\theta)=\int_{\Omega}\hat p(x,y;\theta)\,\dd\Omega.
\label{eq:res_gauge}
\end{equation}
We therefore write the case-wise objective in the unified form
\begin{equation}
\mathcal{L}_{\mathrm{case}}=\lambda_{\mathrm{res}}\|\mathcal{R}_{\Omega}\|_2^2+\lambda_{\mathrm{b}}\|\mathcal{R}_{\partial\Omega}\|_2^2+\lambda_{\mathrm{gauge}}|\mathcal{R}_{\mathrm{g}}|^2,
\label{eq:case_residual}
\end{equation}
and then combine it with data and regularization terms through
\begin{equation}
\mathcal{L}_{\mathrm{total}}=\mathcal{L}_{\mathrm{case}}+\lambda_{\mathrm{data}}\mathcal{L}_{\mathrm{data}}+\lambda_{\mathrm{reg}}\mathcal{L}_{\mathrm{reg}}.
\label{eq:loss_total}
\end{equation}
This notation keeps the solver comparison consistent across all four benchmarks while allowing each case to activate the physics terms that its governing system requires.
The benchmark-specific activation logic is summarized in \SMSec{I}, the composite objective and gauge treatment are detailed in \SMSec{IV}, and the exact metric definitions used in \Tab{tab:error_all} are listed in \SMSec{VI}.

For all four uploaded cases, the benchmark builder prescribes the source terms, coefficients, and boundary data together with reference field data on the evaluation grid. We therefore solve the same PDE template and then compare the learned fields with the provided benchmark maps and the logged quantitative metrics. The field-wise and case-averaged errors appear in \Tab{tab:error_all}, while the wall-clock training-time comparison appears in \Tab{tab:time_all}.

\begingroup
\footnotesize
\setlength{\LTleft}{0pt}
\setlength{\LTright}{0pt}
\begin{longtable}{ccccccc}
\caption{Fieldwise and average error table for all four uploaded benchmark cases. Each model reports the five physical fields $u$, $v$, $p$, $T$, and $\phi$, together with an Avg. row over the five fields. The minimum value for each case, field, and metric appears in bold.}\\
\label{tab:error_all}\\
\toprule
Case & Model & Field & MSE & RMSE & MAE & Relative $L_2$ \\
\midrule
\endfirsthead
\toprule
Case & Model & Field & MSE & RMSE & MAE & Relative $L_2$ \\
\midrule
\endhead
\midrule
\multicolumn{7}{r}{Continued on next page}\\
\midrule
\endfoot
\bottomrule
\endlastfoot
Case 1 & LSTM-PINN & u & $\mathbf{2.675e-7}$ & $\mathbf{5.172e-4}$ & $\mathbf{3.965e-4}$ & $\mathbf{1.368e-3}$ \\
 &  & v & $\mathbf{2.429e-7}$ & $\mathbf{4.929e-4}$ & $\mathbf{3.785e-4}$ & $\mathbf{1.172e-3}$ \\
 &  & p & $\mathbf{1.228e-7}$ & $\mathbf{3.505e-4}$ & $\mathbf{2.416e-4}$ & $\mathbf{3.343e-3}$ \\
 &  & T & $\mathbf{4.024e-7}$ & $\mathbf{6.343e-4}$ & $4.974e-4$ & $\mathbf{1.223e-3}$ \\
 &  & phi & $\mathbf{6.430e-9}$ & $\mathbf{8.019e-5}$ & $\mathbf{6.276e-5}$ & $\mathbf{1.842e-4}$ \\
 &  & Avg. & $\mathbf{2.084e-7}$ & $\mathbf{4.150e-4}$ & $\mathbf{3.153e-4}$ & $\mathbf{1.458e-3}$ \\
\midrule
 & ResAttn-PINN & u & $2.680e-7$ & $5.177e-4$ & $4.040e-4$ & $1.369e-3$ \\
 &  & v & $3.816e-7$ & $6.177e-4$ & $4.856e-4$ & $1.469e-3$ \\
 &  & p & $1.840e-7$ & $4.289e-4$ & $3.070e-4$ & $4.090e-3$ \\
 &  & T & $4.697e-7$ & $6.853e-4$ & $\mathbf{4.928e-4}$ & $1.322e-3$ \\
 &  & phi & $2.465e-8$ & $1.570e-4$ & $1.097e-4$ & $3.607e-4$ \\
 &  & Avg. & $2.656e-7$ & $4.813e-4$ & $3.598e-4$ & $1.722e-3$ \\
\midrule
 & pLSTM-PINN & u & $7.500e-6$ & $2.739e-3$ & $2.187e-3$ & $7.243e-3$ \\
 &  & v & $1.399e-5$ & $3.740e-3$ & $2.939e-3$ & $8.892e-3$ \\
 &  & p & $2.325e-5$ & $4.822e-3$ & $4.151e-3$ & $4.599e-2$ \\
 &  & T & $1.451e-4$ & $1.205e-2$ & $6.630e-3$ & $2.323e-2$ \\
 &  & phi & $1.538e-5$ & $3.922e-3$ & $3.251e-3$ & $9.010e-3$ \\
 &  & Avg. & $4.105e-5$ & $5.454e-3$ & $3.832e-3$ & $1.887e-2$ \\
\midrule
 & Pure-MLP & u & $5.053e-5$ & $7.109e-3$ & $5.503e-3$ & $1.880e-2$ \\
 &  & v & $4.231e-5$ & $6.505e-3$ & $5.302e-3$ & $1.547e-2$ \\
 &  & p & $3.341e-5$ & $5.780e-3$ & $3.932e-3$ & $5.512e-2$ \\
 &  & T & $1.480e-4$ & $1.216e-2$ & $9.407e-3$ & $2.346e-2$ \\
 &  & phi & $4.304e-7$ & $6.560e-4$ & $4.185e-4$ & $1.507e-3$ \\
 &  & Avg. & $5.493e-5$ & $6.443e-3$ & $4.913e-3$ & $2.287e-2$ \\
\midrule
Case 2 & LSTM-PINN & u & $\mathbf{4.068e-7}$ & $\mathbf{6.378e-4}$ & $\mathbf{5.659e-4}$ & $1.041e-2$ \\
 &  & v & $\mathbf{1.631e-7}$ & $\mathbf{4.039e-4}$ & $\mathbf{3.355e-4}$ & $6.863e-3$ \\
 &  & p & $1.572e-5$ & $3.965e-3$ & $3.928e-3$ & $5.137e-2$ \\
 &  & T & $\mathbf{9.236e-8}$ & $\mathbf{3.039e-4}$ & $\mathbf{2.458e-4}$ & $\mathbf{6.307e-4}$ \\
 &  & phi & $\mathbf{8.745e-9}$ & $\mathbf{9.351e-5}$ & $\mathbf{7.858e-5}$ & $\mathbf{2.585e-4}$ \\
 &  & Avg. & $3.279e-6$ & $\mathbf{1.081e-3}$ & $\mathbf{1.031e-3}$ & $\mathbf{1.391e-2}$ \\
\midrule
 & ResAttn-PINN & u & $9.679e-7$ & $9.838e-4$ & $8.440e-4$ & $1.605e-2$ \\
 &  & v & $1.364e-6$ & $1.168e-3$ & $9.485e-4$ & $1.985e-2$ \\
 &  & p & $2.801e-5$ & $5.292e-3$ & $5.269e-3$ & $6.857e-2$ \\
 &  & T & $4.166e-7$ & $6.454e-4$ & $5.323e-4$ & $1.339e-3$ \\
 &  & phi & $1.734e-7$ & $4.164e-4$ & $3.676e-4$ & $1.151e-3$ \\
 &  & Avg. & $6.186e-6$ & $1.701e-3$ & $1.592e-3$ & $2.139e-2$ \\
\midrule
 & pLSTM-PINN & u & $9.593e-6$ & $3.097e-3$ & $2.575e-3$ & $\mathbf{9.123e-3}$ \\
 &  & v & $5.540e-6$ & $2.354e-3$ & $1.863e-3$ & $\mathbf{5.985e-3}$ \\
 &  & p & $\mathbf{3.767e-6}$ & $\mathbf{1.941e-3}$ & $\mathbf{1.511e-3}$ & $\mathbf{1.732e-2}$ \\
 &  & T & $1.254e-4$ & $1.120e-2$ & $9.552e-3$ & $3.056e-2$ \\
 &  & phi & $7.031e-5$ & $8.385e-3$ & $6.650e-3$ & $2.798e-2$ \\
 &  & Avg. & $4.292e-5$ & $5.395e-3$ & $4.430e-3$ & $1.819e-2$ \\
\midrule
 & Pure-MLP & u & $1.705e-6$ & $1.306e-3$ & $1.037e-3$ & $2.131e-2$ \\
 &  & v & $1.246e-6$ & $1.116e-3$ & $8.856e-4$ & $1.897e-2$ \\
 &  & p & $9.544e-6$ & $3.089e-3$ & $3.037e-3$ & $4.003e-2$ \\
 &  & T & $2.894e-6$ & $1.701e-3$ & $1.389e-3$ & $3.531e-3$ \\
 &  & phi & $5.184e-8$ & $2.277e-4$ & $1.744e-4$ & $6.295e-4$ \\
 &  & Avg. & $\mathbf{3.088e-6}$ & $1.488e-3$ & $1.305e-3$ & $1.689e-2$ \\
\midrule
Case 3 & LSTM-PINN & u & $\mathbf{3.609e-7}$ & $\mathbf{6.007e-4}$ & $\mathbf{4.109e-4}$ & $\mathbf{4.508e-4}$ \\
 &  & v & $\mathbf{3.961e-7}$ & $\mathbf{6.294e-4}$ & $\mathbf{4.276e-4}$ & $\mathbf{4.723e-4}$ \\
 &  & p & $7.052e-7$ & $8.398e-4$ & $6.112e-4$ & $1.715e-3$ \\
 &  & T & $8.255e-7$ & $9.086e-4$ & $5.672e-4$ & $1.571e-3$ \\
 &  & phi & $\mathbf{1.076e-8}$ & $\mathbf{1.037e-4}$ & $\mathbf{8.564e-5}$ & $\mathbf{4.033e-4}$ \\
 &  & Avg. & $\mathbf{4.597e-7}$ & $\mathbf{6.164e-4}$ & $\mathbf{4.205e-4}$ & $\mathbf{9.224e-4}$ \\
\midrule
 & ResAttn-PINN & u & $4.705e-7$ & $6.859e-4$ & $5.302e-4$ & $5.147e-4$ \\
 &  & v & $8.408e-7$ & $9.170e-4$ & $6.785e-4$ & $6.881e-4$ \\
 &  & p & $\mathbf{6.388e-7}$ & $\mathbf{7.993e-4}$ & $\mathbf{5.320e-4}$ & $\mathbf{1.632e-3}$ \\
 &  & T & $\mathbf{4.498e-7}$ & $\mathbf{6.707e-4}$ & $\mathbf{4.106e-4}$ & $\mathbf{1.160e-3}$ \\
 &  & phi & $8.873e-8$ & $2.979e-4$ & $2.362e-4$ & $1.158e-3$ \\
 &  & Avg. & $4.977e-7$ & $6.741e-4$ & $4.775e-4$ & $1.031e-3$ \\
\midrule
 & pLSTM-PINN & u & $1.862e-6$ & $1.365e-3$ & $1.079e-3$ & $1.024e-3$ \\
 &  & v & $2.347e-6$ & $1.532e-3$ & $1.278e-3$ & $1.150e-3$ \\
 &  & p & $5.097e-6$ & $2.258e-3$ & $1.798e-3$ & $4.609e-3$ \\
 &  & T & $1.609e-6$ & $1.269e-3$ & $9.732e-4$ & $2.194e-3$ \\
 &  & phi & $5.752e-7$ & $7.584e-4$ & $5.695e-4$ & $2.949e-3$ \\
 &  & Avg. & $2.298e-6$ & $1.436e-3$ & $1.140e-3$ & $2.385e-3$ \\
\midrule
 & Pure-MLP & u & $6.332e-5$ & $7.957e-3$ & $5.886e-3$ & $5.971e-3$ \\
 &  & v & $7.685e-5$ & $8.767e-3$ & $6.447e-3$ & $6.579e-3$ \\
 &  & p & $8.423e-5$ & $9.178e-3$ & $7.144e-3$ & $1.874e-2$ \\
 &  & T & $2.322e-5$ & $4.818e-3$ & $3.559e-3$ & $8.332e-3$ \\
 &  & phi & $5.286e-7$ & $7.271e-4$ & $5.140e-4$ & $2.827e-3$ \\
 &  & Avg. & $4.963e-5$ & $6.289e-3$ & $4.710e-3$ & $8.489e-3$ \\
\midrule
Case 4 & LSTM-PINN & u & $2.349e-9$ & $4.846e-5$ & $3.936e-5$ & $9.893e-5$ \\
 &  & v & $\mathbf{2.128e-10}$ & $\mathbf{1.459e-5}$ & $\mathbf{1.180e-5}$ & $\mathbf{7.148e+8}$ \\
 &  & p & $1.077e-9$ & $3.282e-5$ & $2.603e-5$ & $1.838e-4$ \\
 &  & T & $\mathbf{3.107e-9}$ & $\mathbf{5.574e-5}$ & $\mathbf{4.313e-5}$ & $\mathbf{1.182e-4}$ \\
 &  & phi & $1.215e-10$ & $1.102e-5$ & $\mathbf{7.839e-6}$ & $4.750e-5$ \\
 &  & Avg. & $\mathbf{1.373e-9}$ & $\mathbf{3.253e-5}$ & $\mathbf{2.563e-5}$ & $\mathbf{1.430e+8}$ \\
\midrule
 & ResAttn-PINN & u & $\mathbf{1.149e-9}$ & $\mathbf{3.390e-5}$ & $\mathbf{2.690e-5}$ & $\mathbf{6.920e-5}$ \\
 &  & v & $8.788e-10$ & $2.964e-5$ & $2.288e-5$ & $1.453e+9$ \\
 &  & p & $\mathbf{8.005e-10}$ & $\mathbf{2.829e-5}$ & $\mathbf{2.166e-5}$ & $\mathbf{1.584e-4}$ \\
 &  & T & $5.388e-9$ & $7.340e-5$ & $5.315e-5$ & $1.557e-4$ \\
 &  & phi & $\mathbf{1.160e-10}$ & $\mathbf{1.077e-5}$ & $8.079e-6$ & $\mathbf{4.642e-5}$ \\
 &  & Avg. & $1.666e-9$ & $3.520e-5$ & $2.653e-5$ & $2.905e+8$ \\
\midrule
 & pLSTM-PINN & u & $1.824e-6$ & $1.351e-3$ & $1.136e-3$ & $2.757e-3$ \\
 &  & v & $1.070e-6$ & $1.034e-3$ & $8.820e-4$ & $5.068e+10$ \\
 &  & p & $2.817e-7$ & $5.308e-4$ & $3.870e-4$ & $2.972e-3$ \\
 &  & T & $6.712e-6$ & $2.591e-3$ & $2.299e-3$ & $5.494e-3$ \\
 &  & phi & $7.044e-6$ & $2.654e-3$ & $2.484e-3$ & $1.144e-2$ \\
 &  & Avg. & $3.386e-6$ & $1.632e-3$ & $1.438e-3$ & $1.014e+10$ \\
\midrule
 & Pure-MLP & u & $2.365e-7$ & $4.863e-4$ & $4.024e-4$ & $9.927e-4$ \\
 &  & v & $4.051e-8$ & $2.013e-4$ & $1.643e-4$ & $9.862e+9$ \\
 &  & p & $5.205e-8$ & $2.282e-4$ & $1.761e-4$ & $1.278e-3$ \\
 &  & T & $6.116e-7$ & $7.820e-4$ & $5.881e-4$ & $1.658e-3$ \\
 &  & phi & $2.018e-8$ & $1.421e-4$ & $1.086e-4$ & $6.121e-4$ \\
 &  & Avg. & $1.922e-7$ & $3.680e-4$ & $2.879e-4$ & $1.972e+9$ \\
\end{longtable}
\endgroup

\noindent Table~\ref{tab:error_all} lists the field-wise errors for $u$, $v$, $p$, $\phi$, and $T$ together with the corresponding Avg. rows used for the case-level comparison discussed below. In Case~4, the relative $L_2$ values for the $v$ field and the Avg. row become numerically ill-conditioned because the reference norm is extremely small, so the stable comparison in that case should rely primarily on MSE, RMSE, MAE, and the error maps.
\FloatBarrier

\begin{table}[t]
\centering
\caption{Training time in hours converted from the logged time-cost files.}
\label{tab:time_all}
\small
\setlength{\tabcolsep}{6pt}
\begin{tabular}{cccccc}
\toprule
Case & ResAttn-PINN & LSTM-PINN & pLSTM-PINN & Pure-MLP & Fastest \\
\midrule
Case 1 & 21.91 & 11.00 & 3.51 & 4.31 & pLSTM-PINN \\
Case 2 & 21.22 & 12.28 & 3.73 & 2.27 & Pure-MLP \\
Case 3 & 31.69 & 18.21 & 7.54 & 27.38 & pLSTM-PINN \\
Case 4 & 14.04 & 17.94 & 7.31 & 29.18 & pLSTM-PINN \\
\bottomrule
\end{tabular}
\end{table}
\FloatBarrier

\section{Case 1: Boussinesq Electrothermal Benchmark}

Case 1 investigates the fundamental Boussinesq electrothermal incompressible Navier-Stokes system. We strictly maintain the unified interior residual $\mathcal{N}(\bU)$ defined in \Eq{eq:operator} under no-slip velocity boundaries, mixed thermal and electric constraints, and a zero-mean pressure gauge. This specific configuration establishes the baseline physical regime for the present study. It isolates the core challenge of the problem, ensuring the five-field coupling remains fully active while the constitutive laws remain free from the additional drift or highly non-linear Forchheimer drag introduced in subsequent cases. Consequently, this regime provides a highly transparent environment to evaluate how effectively a neural network manages primary advective and diffusive interactions.

\begin{figure}[t]
    \centering
    \includegraphics[width=0.97\textwidth,height=0.73\textheight,keepaspectratio]{case01.png}
    \caption{Visual comparison of the predicted electrothermal fields for Case 1. The top strip reports the convergence history, while the subsequent rows display the spatial distributions of $u$, $v$, $p$, $\phi$, and $T$. The left block contrasts the exact benchmark against the network predictions, and the right block visualizes the corresponding absolute-error maps.}
    \label{fig:case1}
\end{figure}
\FloatBarrier

Building upon this fundamental coupled setup, the visual evidence in \Fig{fig:case1} demonstrates that the LSTM-PINN framework successfully preserves the physical fidelity of the primary energy-momentum coupling. The memory-enhanced architecture reconstructs all five physical fields with the cleanest global geometry and the weakest background contamination among the evaluated models. Crucially, in high-gradient regions driven by convective transport, the velocity, potential, and temperature channels remain sharply organized around the steep diagonal structures. This structural preservation indicates that the depth-recursive memory mechanism effectively resolves the thin thermal and momentum boundary layers without succumbing to the non-physical boundary layer broadening or visible oscillatory artifacts that afflict the competing models. Furthermore, the corresponding pressure map confirms the capability to maintain smooth macroscopic pressure redistribution while capturing internal localized transitions with minimal thermodynamic distortion. While the ResAttn-PINN formulation produces comparable macroscopic structures, it exhibits slightly higher localized errors, whereas the pLSTM-PINN and Pure-MLP architectures clearly fail to maintain the necessary local sharpness across the convective interfaces.

The quantitative assessment detailed in \Tab{tab:error_all} strongly corroborates these visual observations regarding cross-field consistency. In this Boussinesq regime, the LSTM-PINN framework achieves superior thermodynamic accuracy, securing the lowest error metrics across nearly all individual physical fields and establishing the most robust global performance across mean squared error (MSE), root mean square error (RMSE), mean absolute error (MAE), and relative $L_2$ norms. Specifically, the proposed model achieves an averaged root mean square error of $4.150\times 10^{-4}$, which significantly outperforms the $4.813\times 10^{-4}$ recorded by ResAttn-PINN, the $5.454\times 10^{-3}$ by pLSTM-PINN, and the $6.443\times 10^{-3}$ by Pure-MLP. However, as documented in \Tab{tab:time_all}, this substantial gain in thermodynamic fidelity requires a higher computational investment. The parallelized pLSTM-PINN and the strictly feedforward Pure-MLP converge in 3.51 hours and 4.31 hours respectively, while the depth-recursive LSTM-PINN demands 11.00 hours, remaining significantly more efficient than the 21.91 hours required by the residual-attention network. Ultimately, this evaluation confirms that the depth-recursive memory approach delivers a decisive accuracy advantage and superior structural preservation at a highly justifiable computational cost.

\section{Case 2: Drift-Potential Gauge Benchmark}

Case 2 advances the physical complexity by altering the pressure treatment within the strongly coupled five-field operator. While the continuity, momentum, electric-potential, and temperature formulations remain governed by the interior residuals defined in \Eq{eq:operator}, the absolute pressure is no longer secured by localized point-wise anchoring. Instead, the pressure field must be reconstructed through a global zero-mean gauge constraint associated with \Eq{eq:res_gauge}. Concurrently, the electric-potential equation introduces drift-sensitive coupling into the transport mechanisms. This specific configuration forces the computational framework to reconstruct the macroscopic pressure distribution indirectly while strictly preserving the delicate cross-field consistency required by the interactive electrothermal system.

\begin{figure}[t]
    \centering
    \includegraphics[width=0.97\textwidth,height=0.73\textheight,keepaspectratio]{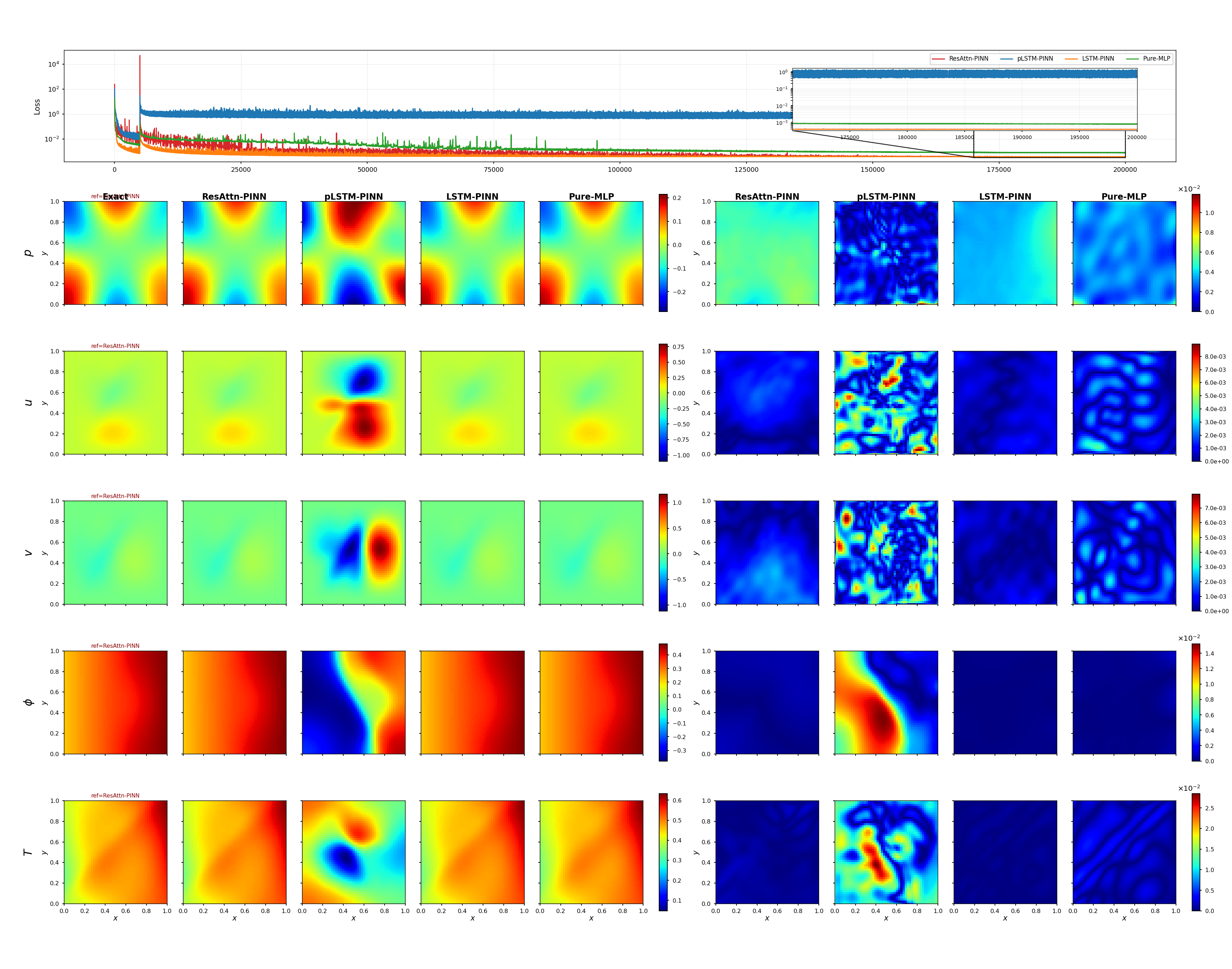}
    \caption{Visual comparison of the predicted fields for the drift-potential gauge configuration (Case 2). This scenario replaces direct pressure anchoring with a global zero-mean gauge constraint and introduces drift-potential coupling, specifically evaluating the capacity of each network to reconstruct the interactive multiphysics fields under indirect macroscopic pressure constraints.}
    \label{fig:case2}
\end{figure}
\FloatBarrier

As visualized in \Fig{fig:case2}, the introduction of the global gauge treatment significantly amplifies the reconstruction difficulty for all evaluated architectures. This macroscopic constraint fundamentally disrupts the local point-wise equilibrium, forcing the fluid momentum to respond to asynchronous body forces without a rigid spatial anchor. Even under this stringent physical setting, the LSTM-PINN framework preserves the most coherent global structural organization across the velocity, temperature, and electric-potential distributions. The pressure reconstruction remains inherently challenging for every solver, and the spatially decoupled pLSTM-PINN actually resolves the specific pressure channel more accurately than the competing backbones. However, this localized advantage reveals a severe physical limitation. The pLSTM-PINN achieves this pressure accuracy by exclusively prioritizing homogeneous pressure redistribution at the expense of multi-field interaction fidelity, resulting in substantial physical distortions within the thermal and electrical potential fields. Conversely, the spatial memory of the LSTM-PINN intrinsically coordinates the localized drift-potential transport with the macroscopic pressure distribution. It accepts extremely minor accuracy compromises in the homogeneous pressure baseline to strictly enforce the global thermodynamic and electrohydrodynamic balance across the remaining interactive fields.

The quantitative assessment presented in \Tab{tab:error_all} explicitly confirms this crucial physical tradeoff. In this drift-potential regime, the LSTM-PINN achieves superior accuracy in the primary interactive fields, securing the lowest error metrics across the velocity, temperature, and electrical potential components, while the pLSTM-PINN yields the lowest absolute errors exclusively in the pressure channel. When evaluating the comprehensive system fidelity, the depth-recursive LSTM-PINN delivers the best overall structural preservation, achieving the lowest global root mean square error, mean absolute error, and relative $L_2$ norm. Specifically, the averaged root mean square error drops to $1.081 \times 10^{-3}$ for the LSTM-PINN, compared with $1.701 \times 10^{-3}$ for ResAttn-PINN, $5.395 \times 10^{-3}$ for pLSTM-PINN, and $1.488 \times 10^{-3}$ for Pure-MLP. Furthermore, \Tab{tab:time_all} indicates that this robust cross-field consistency is achieved at a manageable computational cost. The LSTM-PINN framework requires 12.28 hours for convergence, positioning it as a highly balanced option between the rapid but physically distorted Pure-MLP (2.27 hours) and the substantially slower ResAttn-PINN (21.22 hours). Therefore, the memory-enhanced architecture demonstrates a superior capacity to navigate conflicting physical constraints while maintaining global system fidelity.

\section{Case 3: Buoyancy-Coupled Electrothermal Benchmark}

Case 3 activates the buoyancy-driven electrothermal coupling within the vertical momentum balance, serving as a critical test for resolving strongly non-linear feedback loops. In this physical regime, the residual construction becomes significantly more tightly coupled than in the baseline configuration because local temperature variations directly dictate density gradients and buoyancy forces. These forces subsequently accelerate the fluid and fundamentally alter the convective heat transfer pathways. We maintain the identical state variable set $\{u, v, p, \phi, T\}$ and the same boundary-consistency philosophy to isolate this specific phenomenon. Consequently, the primary computational challenge arises from maintaining multi-field interaction fidelity under this strong thermal-to-momentum feedback, an environment where traditional point-wise neural networks frequently exhibit severe numerical instability.

\begin{figure}[t]
    \centering
    \includegraphics[width=0.97\textwidth,height=0.73\textheight,keepaspectratio]{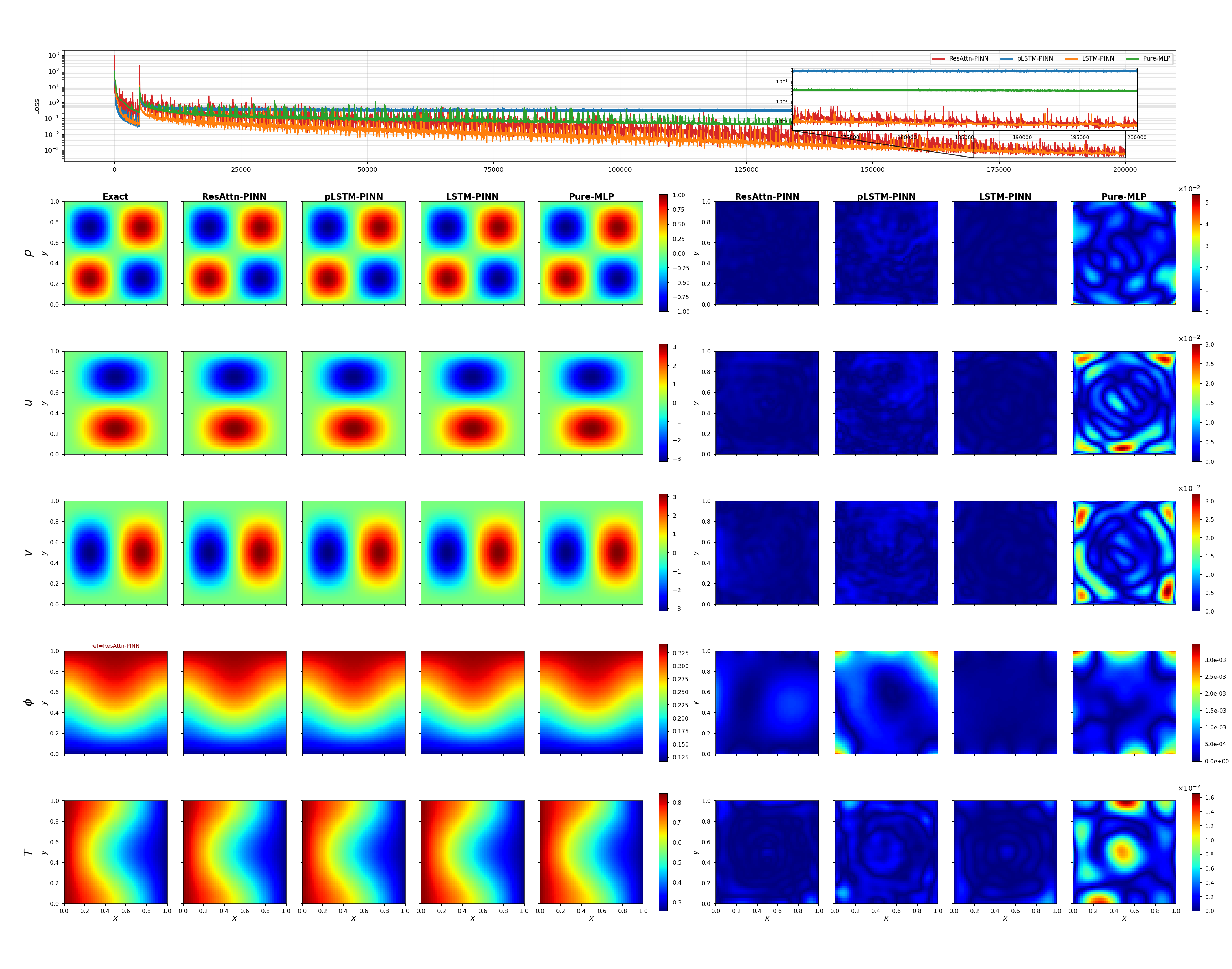}
    \caption{Visual comparison of the predicted fields for the buoyancy-coupled electrothermal configuration (Case 3). This physical regime emphasizes advection-dominated transport and strong thermal-to-momentum feedback, explicitly evaluating the capacity of each architecture to resolve complex convective interactions without inducing numerical artifacts.}
    \label{fig:case3}
\end{figure}
\FloatBarrier

The visualization provided in \Fig{fig:case3} demonstrates a remarkably clear separation between the computational solvers in handling this intense non-linear convection. The LSTM-PINN framework uniquely maintains the authentic physical morphology of the thermal plumes and the associated convective velocity fields. By keeping the advection-dominated field geometry strictly intact, the memory-enhanced architecture accurately captures the momentum and electric potential distributions without introducing the non-physical strip-like artifacts symptomatic of numerical instability within the feedback loop. While the ResAttn-PINN formulation remains highly competitive and achieves marginally sharper spatial resolutions in the scalar pressure and temperature channels, it still permits subtle structural deviations in the broader flow geometry. Conversely, the purely feedforward Pure-MLP merely reproduces the macroscopic thermodynamic tendencies and critically fails to capture the fine-scale boundary layer behaviors. Notably, the spatially decoupled pLSTM-PINN deteriorates most severely in this strongly coupled regime, generating massive structural distortions that completely misrepresent the localized convective heat transfer mechanisms.

The quantitative metrics detailed in \Tab{tab:error_all} substantiate these visual observations and highlight the complex physical trade-offs inherent in strongly coupled multiphysics simulations. In this buoyancy-driven scenario, the LSTM-PINN secures the highest accuracy in resolving the highly dynamic velocity and electrical potential fields, whereas the residual-attention network demonstrates a slight numerical advantage in the scalar pressure and temperature distributions. Despite this localized performance split, the depth-recursive memory approach delivers the most stable macroscopic thermodynamic balance, achieving the lowest global error values across mean squared error, root mean square error, mean absolute error, and relative $L_2$ metrics. Specifically, the globally averaged root mean square error is stabilized at $6.164 \times 10^{-4}$ for the LSTM-PINN, closely followed by $6.741 \times 10^{-4}$ for the ResAttn-PINN, but significantly outperforming the $1.436 \times 10^{-3}$ and $6.289 \times 10^{-3}$ recorded by the pLSTM-PINN and Pure-MLP architectures, respectively. Furthermore, the computational runtime analysis in \Tab{tab:time_all} confirms that the depth-recursive architecture achieves this superior cross-field stability highly efficiently. While it inherently demands more computational resources than the heavily distorted decoupled models, the LSTM-PINN framework reaches convergence in substantially less time than both the residual-attention network and the dense multilayer perceptron, establishing an exceptionally favorable balance between thermodynamic accuracy and computational efficiency.

\section{Case 4: Brinkman-Forchheimer Drift Benchmark}

Case 4 introduces the Brinkman-Forchheimer drag formulation coupled with drift-potential transport, establishing the most physically complex regime in this study. The introduction of this non-linear drag fundamentally breaks the linear fluid shear equilibrium and imposes intense, localized dissipation at the macroscopic interfaces. Consequently, the computational solver must concurrently coordinate macroscopic boundary constraints, micro-scale non-linear drag forces, and the intricate multi-field coupling among fluid flow, electric potential, and thermal energy. Because this specific physical configuration inherently restricts the absolute magnitude of the thermodynamic variations, generating the smallest absolute errors across all evaluated regimes, the comparative analysis becomes exceptionally sensitive to subtle structural deviations in the physical fields.

\begin{figure}[t]
    \centering
    \includegraphics[width=0.97\textwidth,height=0.73\textheight,keepaspectratio]{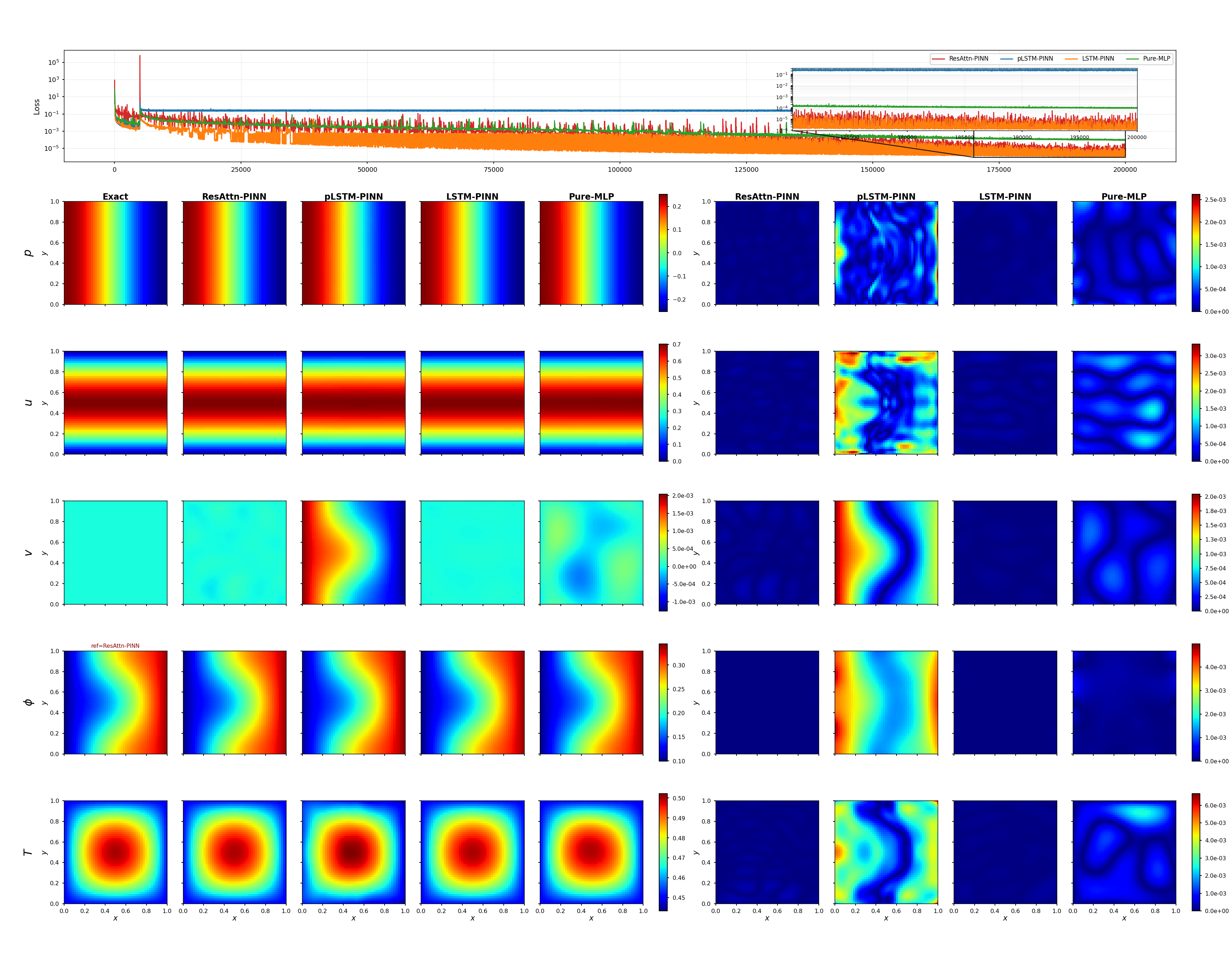}
    \caption{Visual comparison of the predicted fields for the Brinkman-Forchheimer drift configuration (Case 4). This regime combines intense non-linear interfacial drag with drift-potential coupling, specifically highlighting the rigorous competition between the memory-enhanced and residual-attention architectures by displaying the absolute-error maps alongside the reconstructed multiphysics distributions.}
    \label{fig:case4}
\end{figure}
\FloatBarrier

As depicted in \Fig{fig:case4}, both the LSTM-PINN and the ResAttn-PINN successfully navigate this severe non-linear dissipation, resolving the electrothermal benchmark with remarkably low absolute errors. The depth-recursive memory of the LSTM-PINN framework excels in preserving clean convection structures in the vertical momentum ($v$) and thermal ($T$) channels without inducing numerical oscillations. Conversely, the ResAttn-PINN achieves slightly sharper gradient resolutions in the horizontal velocity ($u$), pressure ($p$), and electric potential ($\phi$) distributions. In stark contrast, the spatially decoupled pLSTM-PINN struggles with the asynchronous body forces, generating large drift-aligned non-physical oscillations. Furthermore, the Pure-MLP introduces broad structural distortions across the entire computational domain, failing to capture the delicate interfacial momentum balances.

The quantitative data in \Tab{tab:error_all} explicitly detail this highly competitive multi-field resolution. The individual field metrics indicate a nuanced split, where the ResAttn-PINN demonstrates a slight numerical edge in capturing the $u$, $p$, and $\phi$ components, while the LSTM-PINN maintains superior fidelity in resolving the $v$ and $T$ interactions. Despite this localized competition, the depth-recursive architecture continuously coordinates the macroscopic pressure drop with the micro-scale non-linear drag forces, ultimately retaining the most balanced global thermodynamic consistency across the mean squared error, root mean square error, and mean absolute error. Specifically, the globally averaged root mean square error registers at an exceptionally low $3.253 \times 10^{-5}$ for the LSTM-PINN, narrowly outperforming the $3.520 \times 10^{-5}$ achieved by the ResAttn-PINN. It is important to note that the relative $L_2$ error metric becomes numerically ill-conditioned in this interface-dominated regime because the physical reference norm of the transverse velocity field approaches zero, causing the relative ratio to mathematically inflate despite the extremely low absolute discrepancies. Furthermore, the computational efficiency analysis in \Tab{tab:time_all} reflects the rigorous demands of resolving this intense non-linear dissipation. The LSTM-PINN framework requires 17.94 hours for convergence, which is longer than both the pLSTM-PINN (7.31 hours) and the ResAttn-PINN (14.04 hours). Consequently, this challenging regime confirms that the memory-enhanced architecture maintains a robust, albeit modest, global thermodynamic advantage under severe non-linear drag, validating its structural reliability despite the increased computational investment.

\section*{Supplementary Material}
Supplementary material accompanies this article. It now contains the extended technical support items omitted from the main text: notation and case activation in \SMSec{I}, recovered architecture of all four compared backbones in \SMSec{II}, full case-wise hyperparameter tables for every network in \SMSec{III}, composite objective and gauge treatment in \SMSec{IV}, benchmark protocol and archive bookkeeping in \SMSec{V}, metric definitions and runtime accounting in \SMSec{VI}, supporting literature inventory in \SMSec{VII}, and manuscript--supplement correspondence in \SMSec{VIII}.

\section{Conclusion}
This study demonstrates that the LSTM-PINN framework provides a robust and highly accurate computational approach for resolving steady-state electrothermal systems within a unified multiphysics formulation. By systematically evaluating the depth-recursive architecture across diverse physical scenarios, we establish its superior capability to maintain strict cross-field consistency and thermodynamic fidelity compared to conventional and attention-based networks. The physical advantages of this memory-enhanced approach are particularly evident in the Boussinesq and strong buoyancy-coupled regimes, where the network successfully captures complex thermal convection structures and decisively suppresses non-physical structural artifacts. Furthermore, under stringent global pressure-gauge constraints and highly non-linear Brinkman-Forchheimer drag interactions, the framework continues to deliver the optimal balance of absolute accuracy across all five coupled physical fields. Although achieving this stringent level of structural preservation requires a higher computational training cost than purely feedforward models, the significant gains in predictive reliability firmly justify the methodology. Ultimately, this memory-aware architecture establishes a powerful and reliable baseline for accurately simulating complex momentum, thermal energy, and electric transport phenomena in advanced multiphysics applications.

\printcredits

\section*{Declaration of competing interest}
The authors declared that they have no conflicts of interest to this work.

\section*{Acknowledgment}
This work is supported by the Developing Project of Science and Technology of Jilin Province (20250102032JC).

\section*{Data availability}
All the code for this article is available open access at a Github repository available at https://github.com/Uderwood-TZ/LSTM-PINN-for-Solving-Steady-State-Electrothermal-Coupled-Multiphysics-Problems.git.
\bibliographystyle{unsrtnat}
\bibliography{references}

@article{heatreview2025,
  author  = {Zhuang Zhao and Ye Wang and Weijian Zhang and Zhenggang Ba and Lin Sun},
  title   = {Physics-informed neural networks in heat transfer-dominated multiphysics systems: A comprehensive review},
  journal = {Engineering Applications of Artificial Intelligence},
  year    = {2025},
  volume  = {157},
  pages   = {111098},
  doi     = {10.1016/j.engappai.2025.111098}
}

@article{borehole2025,
  author  = {Pengchao Li and Fang Guo and Yongfei Li and Xuejing Yang and Xudong Yang},
  title   = {Physics-informed neural network for real-time thermal modeling of large-scale borehole thermal energy storage systems},
  journal = {Energy},
  year    = {2025},
  volume  = {315},
  pages   = {134344},
  doi     = {10.1016/j.energy.2024.134344}
}

@article{thermoelectric2025,
  author  = {Hyeonbin Moon and others},
  title   = {Physics-informed neural operators for generalizable and label-free inference of temperature-dependent thermoelectric properties},
  journal = {npj Computational Materials},
  year    = {2025},
  volume  = {11},
  pages   = {272},
  doi     = {10.1038/s41524-025-01769-1}
}

@article{heatstorage2025,
  author  = {Gwangwoo Han and Sung Kook Hong and Joo Hyun Moon},
  title   = {Physics-informed neural network digital twin for real-time energy-state tracking in multivariate thermochemical heat storage},
  journal = {Journal of Energy Storage},
  year    = {2025},
  volume  = {139},
  pages   = {118686},
  doi     = {10.1016/j.est.2025.118686}
}

@article{pisanet2026,
  author  = {Chaobo Zhang and others},
  title   = {{PISA-Net}: A physics-informed structure-aware neural network for multiphysics field reconstruction in liquid cooling systems},
  journal = {Applied Thermal Engineering},
  year    = {2026},
  doi     = {10.1016/j.applthermaleng.2025.129347},
  note    = {Online 2025}
}

@article{batteryhybrid2025,
  author  = {Yufu Luo and others},
  title   = {A method for estimating lithium-ion battery state of health based on physics-informed hybrid neural network},
  journal = {Electrochimica Acta},
  year    = {2025},
  volume  = {525},
  pages   = {146110},
  doi     = {10.1016/j.electacta.2025.146110}
}

@article{batterypiml2026,
  author  = {Runze Wang and others},
  title   = {Physics-informed machine learning method for lithium-ion batteries state of health estimation using simulated voltage data},
  journal = {Applied Energy},
  year    = {2026},
  note    = {In press}
}

@article{batterycoest2025,
  author  = {Li Yang and Mingjian He and Yatao Ren and Baohai Gao and Hong Qi},
  title   = {Physics-informed neural network for co-estimation of state of health, remaining useful life, and short-term degradation path in lithium-ion batteries},
  journal = {Applied Energy},
  year    = {2025},
  volume  = {398},
  pages   = {126427},
  doi     = {10.1016/j.apenergy.2025.126427}
}

@article{batterysoc2025,
  author  = {Donghee Son and Shina Park and Junseok Oh and Taehan Lee and Sang Woo Kim},
  title   = {Physics-informed dual-stage network for lithium-ion battery state-of-charge estimation under various aging and temperature conditions},
  journal = {Applied Energy},
  year    = {2025},
  volume  = {401},
  pages   = {126770},
  doi     = {10.1016/j.apenergy.2025.126770}
}

@article{pemfc2026,
  author  = {Julian Nicolas Toussaint and Sebastian Pieper and Max Paul Mally and Stefan Pischinger},
  title   = {Leveraging the capabilities of physics-informed neural networks for channel optimization in {PEM} fuel cells},
  journal = {Applied Energy},
  year    = {2026},
  volume  = {409},
  pages   = {127478},
  doi     = {10.1016/j.apenergy.2026.127478}
}

@article{jahanbakhsh2024thinfilm,
  author  = {Amirmohammad Jahanbakhsh and Rojan Firuznia and Sina Nazifi and Hadi Ghasemi},
  title   = {Physics-informed neural network for thin-film evaporation in hierarchical structures},
  journal = {International Journal of Heat and Mass Transfer},
  year    = {2024},
  volume  = {224},
  pages   = {125296},
  doi     = {10.1016/j.ijheatmasstransfer.2024.125296}
}

@article{zheng2024tph,
  author  = {Jinglai Zheng and Fan Li and Haiming Huang},
  title   = {{T-phPINN}: Physics-informed neural networks for solving 2D non-Fourier heat conduction equations},
  journal = {International Journal of Heat and Mass Transfer},
  year    = {2024},
  volume  = {235},
  pages   = {126216},
  doi     = {10.1016/j.ijheatmasstransfer.2024.126216}
}

@article{jalili2024bubble,
  author  = {Darioush Jalili and Mohammad Jadidi and Amir Keshmiri and Bhaskar Chakraborty and Anastasios Georgoulas and Yasser Mahmoudi},
  title   = {Transfer learning through physics-informed neural networks for bubble growth in superheated liquid domains},
  journal = {International Journal of Heat and Mass Transfer},
  year    = {2024},
  volume  = {232},
  pages   = {125940},
  doi     = {10.1016/j.ijheatmasstransfer.2024.125940}
}

@article{wang2025packedbed,
  author  = {Mouhao Wang and Shanshan Bu and Bing Zhou and Baoping Gong and Zhenzhong Li and Deqi Chen},
  title   = {Physics-informed neural networks for effective diffusion characteristics inversion in packed bed at low flow rates},
  journal = {International Journal of Heat and Mass Transfer},
  year    = {2025},
  volume  = {244},
  pages   = {126970},
  doi     = {10.1016/j.ijheatmasstransfer.2025.126970}
}

@article{jalili2025filmboiling,
  author  = {Darioush Jalili and Yasser Mahmoudi},
  title   = {Physics-informed neural networks for two-phase film boiling heat transfer},
  journal = {International Journal of Heat and Mass Transfer},
  year    = {2025},
  volume  = {241},
  pages   = {126680},
  doi     = {10.1016/j.ijheatmasstransfer.2025.126680}
}

@article{tang2025thermalwave,
  author  = {Hong Tang and Alexander Melnikov and MingRui Liu and Stefano Sfarra and Hai Zhang and Andreas Mandelis},
  title   = {Physics informed neural networks for solving inverse thermal wave coupled boundary-value problems},
  journal = {International Journal of Heat and Mass Transfer},
  year    = {2025},
  volume  = {245},
  pages   = {126985},
  doi     = {10.1016/j.ijheatmasstransfer.2025.126985}
}

@article{cai2025surfaceflux,
  author  = {Zemin Cai and Xiangqi Lin and Tianshu Liu},
  title   = {Physics-informed neural network for estimating surface heat flux from surface temperature measurements},
  journal = {International Journal of Heat and Mass Transfer},
  year    = {2025},
  volume  = {253},
  pages   = {127580},
  doi     = {10.1016/j.ijheatmasstransfer.2025.127580}
}

@article{gao2025rbf,
  author  = {Kangcheng Gao and Lin Qiu and Gefei Li and Fajie Wang and Ji Lin and Yan Gu},
  title   = {A physics-informed framework based on radial basis functions for forward and inverse heat conduction problems in anisotropic materials},
  journal = {International Journal of Heat and Mass Transfer},
  year    = {2025},
  volume  = {252},
  pages   = {127517},
  doi     = {10.1016/j.ijheatmasstransfer.2025.127517}
}

@article{rba2024,
  author  = {S. J. Anagnostopoulos and others},
  title   = {Residual-based attention in physics-informed neural networks},
  journal = {Computer Methods in Applied Mechanics and Engineering},
  year    = {2024}
}

@article{lapinn2024,
  author  = {Yuan Song and others},
  title   = {Loss-attentional physics-informed neural networks},
  journal = {Journal of Computational Physics},
  year    = {2024}
}

@article{balanced2025,
  author  = {Wen Chen and others},
  title   = {Self-adaptive weights based on balanced residual decay rates in physics-informed neural networks},
  journal = {Journal of Computational Physics},
  year    = {2025}
}

@article{adual2025,
  author  = {Ting Wang and others},
  title   = {Adaptive deep physics-informed neural network with dual adaptivity for complex partial differential equations},
  journal = {Computer Methods in Applied Mechanics and Engineering},
  year    = {2025}
}

@article{ferrer2024gapinn,
  author  = {Antonio Ferrer-S{\'a}nchez and Jos{\'e} D. Mart{\'i}n-Guerrero and Roberto Ruiz de Austri-Bazan and Alejandro Torres-Forn{\'e} and Jos{\'e} A. Font},
  title   = {Gradient-annihilated {PINNs} for solving {Riemann} problems: Application to relativistic hydrodynamics},
  journal = {Computer Methods in Applied Mechanics and Engineering},
  year    = {2024},
  volume  = {424},
  pages   = {116906},
  doi     = {10.1016/j.cma.2024.116906}
}

@article{mdpinn2025,
  author  = {Ting Wang and others},
  title   = {Mixed-depth physics-informed neural network with nested finite basis functions for complex partial differential equations},
  journal = {Computer Methods in Applied Mechanics and Engineering},
  year    = {2025}
}

@article{ddd2024,
  author  = {Victorita Dolean and Alexander Heinlein and Siddhartha Mishra and Ben Moseley},
  title   = {Multilevel domain decomposition-based architectures for physics-informed neural networks},
  journal = {Computer Methods in Applied Mechanics and Engineering},
  year    = {2024},
  pages   = {117116},
  doi     = {10.1016/j.cma.2024.117116}
}

@article{afeinn2025,
  author  = {Santiago Badia and Wei Li and Alberto F. Mart{\'i}n},
  title   = {Adaptive finite element interpolated neural networks},
  journal = {Computer Methods in Applied Mechanics and Engineering},
  year    = {2025}
}

@article{cfeinn2025,
  author  = {Santiago Badia and Wei Li and Alberto F. Mart{\'i}n},
  title   = {Compatible finite element interpolated neural networks},
  journal = {Computer Methods in Applied Mechanics and Engineering},
  year    = {2025},
  pages   = {117889},
  doi     = {10.1016/j.cma.2025.117889}
}

@article{multiscale2024,
  author  = {Y. Wang and others},
  title   = {A practical {PINN} framework for multi-scale problems with reconstructing the loss function and associating it with specialized neural network architectures},
  journal = {Journal of Computational Physics},
  year    = {2024}
}

@article{naspinn2024,
  author  = {Y. Wang and others},
  title   = {{NAS-PINN}: Neural architecture search-guided physics-informed neural network for solving partial differential equations},
  journal = {Journal of Computational Physics},
  year    = {2024}
}

@article{laf2025,
  author  = {Afrah Farea and Mustafa Serdar Celebi},
  title   = {Learnable activation functions in physics-informed neural networks for solving partial differential equations},
  journal = {Computer Physics Communications},
  year    = {2025},
  pages   = {109753}
}

@article{tsa2025,
  author  = {A. Khademi and others},
  title   = {Physics-informed neural networks with trainable sinusoidal activation for solving {Navier--Stokes} equations},
  journal = {Computer Physics Communications},
  year    = {2025}
}

@article{pinto2025,
  author  = {Sumanth Kumar Boya and Deepak N. Subramani},
  title   = {{PINTO}: Physics-informed transformer neural operator for learning generalized solutions of partial differential equations for any initial and boundary condition},
  journal = {Computer Physics Communications},
  year    = {2025},
  pages   = {109702},
  doi     = {10.1016/j.cpc.2025.109702}
}

@article{pinoinfiltration2025,
  author  = {Hamza Kamil and Azzeddine Soula{\"i}mani and Abdelaziz Beljadid},
  title   = {Physics-informed neural operators for efficient modeling of infiltration in porous media},
  journal = {Journal of Computational Physics},
  year    = {2025},
  volume  = {538},
  pages   = {114156},
  doi     = {10.1016/j.jcp.2025.114156}
}

@article{curse2024,
  author  = {Z. Hu and others},
  title   = {Tackling the curse of dimensionality with physics-informed neural networks},
  journal = {Neural Networks},
  year    = {2024}
}

@article{tao2025lstm,
  title={LSTM-PINN: An Hybrid Method for Prediction of Steady-State Electrohydrodynamic Flow},
  author={Tao, Ze and Xu, Ke and Liu, Fujun},
  journal={Journal of Computational Physics},
  pages={114586},
  year={2025},
  publisher={Elsevier}
}

@article{buildingct2026,
  author  = {Dan Wang and Wenju Hu and Wei Wang},
  title   = {Physics-informed neural networks with continuous-time thermal modelling for model predictive control of building energy systems in demand response},
  journal = {Energy and Buildings},
  year    = {2026},
  volume  = {352},
  pages   = {116864}
}

@article{xing2025modeling,
  title   = {Modeling dynamic gas--liquid interfaces in underwater explosions using interval-constrained physics-informed neural networks},
  author  = {Xing, Fulin and Li, Junjie and Tao, Ze and Liu, Fujun and Tan, Yong},
  journal = {Physics of Fluids},
  volume  = {37},
  number  = {11},
  year    = {2025},
  publisher = {AIP Publishing}
}

@article{tao2025analytical,
  title   = {Analytical and neural network approaches for solving two-dimensional nonlinear transient heat conduction},
  author  = {Tao, Ze and Liu, Fujun and Li, Jinhua and Chen, Guibo},
  journal = {arXiv preprint arXiv:2504.02845},
  year    = {2025}
}

@article{tao2025lstm_population,
  title   = {An LSTM-PINN Hybrid Method to the specific problem of population forecasting},
  author  = {Tao, Ze},
  journal = {arXiv preprint arXiv:2505.01819},
  year    = {2025}
}

@article{tao2025lnn,
  title   = {{LNN-PINN}: A unified physics-only training framework with liquid residual blocks},
  author  = {Tao, Ze and Wang, Hanxuan and Liu, Fujun},
  journal = {arXiv preprint arXiv:2508.08935},
  year    = {2025}
}

@article{zhou2026residual,
  title   = {Residual Attention Physics-Informed Neural Networks for Robust Multiphysics Simulation of Steady-State Electrothermal Energy Systems},
  author  = {Zhou, Yuqing and Tao, Ze and Liu, Fujun},
  journal = {arXiv preprint arXiv:2603.23578},
  year    = {2026}
}

\end{document}